\documentclass[aps,nofootinbib,preprintnumbers,superscriptaddress,twocolumn,amsmath,amssymb,floatfix,prl]{revtex4-1} 
\pdfoutput=1
\usepackage{graphicx}
\usepackage[usenames,dvipsnames]{xcolor}
\usepackage[colorlinks,bookmarks=false,citecolor=NavyBlue,linkcolor=Red,urlcolor=blue]{hyperref}
\usepackage{dsfont}
\usepackage[export]{adjustbox}
\usepackage{bbold}
\usepackage{xcolor}

\newcommand\be{\begin{equation}}
\newcommand\bea{\begin{eqnarray}}
\newcommand\bes{\begin{subequations}}
\newcommand\esu{\end{subequations}}
\newcommand\ee{\end{equation}}
\newcommand\eea{\end{eqnarray}}

\newcommand{\cmmnt}[1]{}

\def\doi{http://dx.doi.org/}

\newcommand{\upa}{|\!\!\uparrow\rangle}
\newcommand{\dna}{|\!\!\downarrow\rangle}

\usepackage{braket}

\begin{document}

\title{Suppression of transport in non-disordered quantum spin chains \\ due to confined excitations}

\author{Paolo Pietro Mazza}
\affiliation{SISSA -- International School for Advanced Studies \& INFN, via Bonomea 265, 34136 Trieste, Italy}

\author{Gabriele Perfetto}
\affiliation{SISSA -- International School for Advanced Studies \& INFN, via Bonomea 265, 34136 Trieste, Italy}

\author{Alessio Lerose}
\affiliation{SISSA -- International School for Advanced Studies \& INFN, via Bonomea 265, 34136 Trieste, Italy}

\author{Mario Collura}
\affiliation{The Rudolf Peierls Centre for Theoretical Physics, Oxford University, Oxford, OX1 3NP, United Kingdom}

\author{Andrea Gambassi}
\affiliation{SISSA -- International School for Advanced Studies \& INFN, via Bonomea 265, 34136 Trieste, Italy}


\begin{abstract}
The laws of thermodynamics require any initial macroscopic inhomogeneity in extended many-body systems  to be smoothed out by the time evolution through the activation of transport processes. 
In generic quantum systems, transport is expected to be governed by a diffusion law, whereas a sufficiently strong quenched disorder can suppress it completely due to many-body localization of quantum excitations. 
Here we show that the confinement of quasi-particles can also suppress transport even if the dynamics are generated by non-disordered Hamiltonians. 
We demonstrate this in the quantum Ising chain with transverse and longitudinal magnetic fields, prepared in a paradigmatic state with a domain-wall and therefore with a spatially varying energy density.
We perform extensive numerical simulations of the dynamics which turn out to be in excellent agreement with an effective analytical description valid within both weak and strong confinement regimes. 
Our results show that the energy flow from ``hot'' to ``cold''  regions of the chain is suppressed for all accessible times. We argue that this phenomenon is general, as it relies solely on the emergence of confinement of excitations.
\end{abstract}


\maketitle

\emph{Introduction.---}%
Transport is the fundamental mechanism which allows both classical and quantum isolated and extended statistical systems to smooth out any inhomogeneity possibly present in their initial conditions, while relaxing towards their stationary states. 
In the quantum realm, the interest in this aspect of non-equilibrium dynamics \cite{Noneq, Noneq1, Noneq2} has been recently prompted by an impressive advance in experimental techniques with cold atoms which made it possible to maintain coherent quantum dynamics for sufficiently long times \cite{exp1, exp2, exp3, exp4, exp5, exp6, exp7, exp8, exp9, exp10, exp11, exp12, exp13, exp14, exp15}. 
%
%
 
Of paramount importance, in this context, is to understand whether and how the transport of conserved physical quantities, such as particle and energy densities, occurs~\cite{transexp1,transexp2}.
Generically, the spatial spreading of local inhomogeneities is expected to obey 
a diffusion law, whose microscopic origin is usually traced back to inelastic collisions \cite{mesoscopic1, mesoscopic2}. 
In the specific case of integrable systems, instead --- characterized by an infinite set of (quasi-)local conserved quantities --- transport is enhanced by the existence of stable excitations traveling ballistically with certain characteristic velocities, typically exposed after a sudden change (\emph{quench}) in the parameters of the systems \cite{quench1, quench2, Igloi1, Igloi2, quench3}. 
Correspondingly, a non-equilibrium stationary state may arise, supporting ballistic transport and thus finite currents 
at long times~\cite{cft1,
 trans2,trans3,trans4,
 trans7,
 trans10,trans11,trans12,trans13,ghd1,ghd2
}.  
A completely different scenario emerges in the presence of strong disorder. In fact, in the so-called many-body localized phase \cite{anderson,basko,huse}, the localization of excitations \cite{husephen,ros,abanin} suppresses the energy and particle transport and the system fails to thermalize: initial gradients of local quantities persist for arbitrarily long times during the evolution \cite{vipin1,vipin2,Moore}. 

Disordered-induced localization, however, is not the sole mechanism which hampers the propagation of information in many-body interacting systems \cite{ProsenMBL,Bardarson,YaoTranslInv,Jamir}. 
Indeed, even in the absence of disorder, the dynamical \emph{confinement} of excitations \cite{McCoy,ShankarConfinement} can suppress the spreading of correlations \cite{Calabrese}.
How can this be reconciled with the heuristic expectation that an initial inhomogeneous configuration has to be smoothed out by the evolution?  
In this work we address this issue and show that a non-integrable, non-disordered quantum spin chain with confined excitations, initially prepared in domain-wall states causing a finite energy gradient across the system, can exhibit suppression of energy transport.
 Although we focus here on the paradigmatic quantum Ising chain, introduced further below, the results we obtain are general and apply to a variety of physical systems, as they solely rely on the confinement of the excitations, which has been shown to emerge in several other one-dimensional condensed-matter models~\cite{SchollwockRealTimeConfinement,WangConfinement,EsslerConfinement,Rut2,GorshkovConfinement,GiappiHiggs,S1Confinement1} as well as lattice gauge theories~\cite{Montangero,CiracDemlerVariational,SuraceRydberg}.

\begin{figure*}[!t]
\includegraphics[scale=0.45]{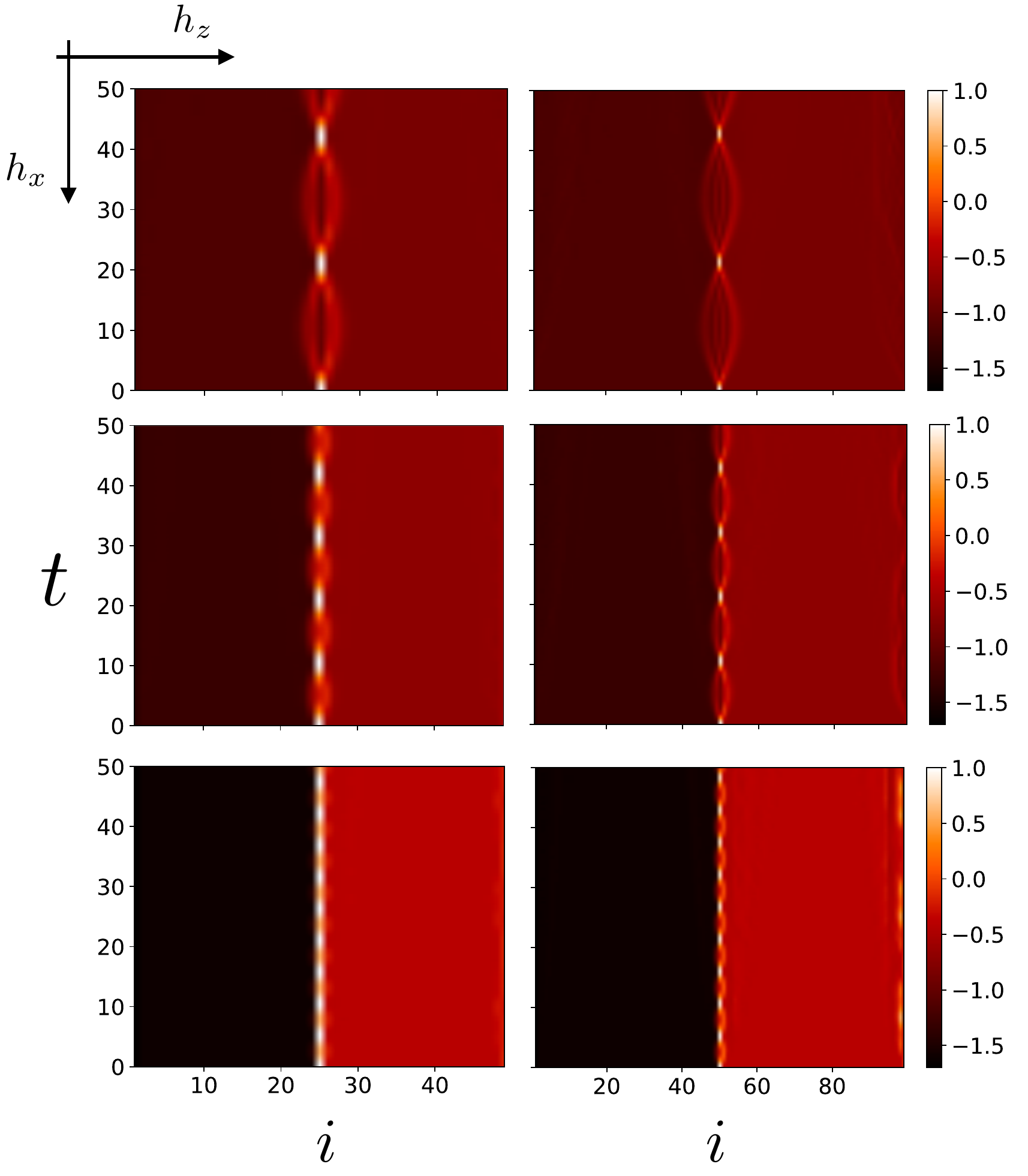}\hspace{2pc}
\includegraphics[scale=0.45]{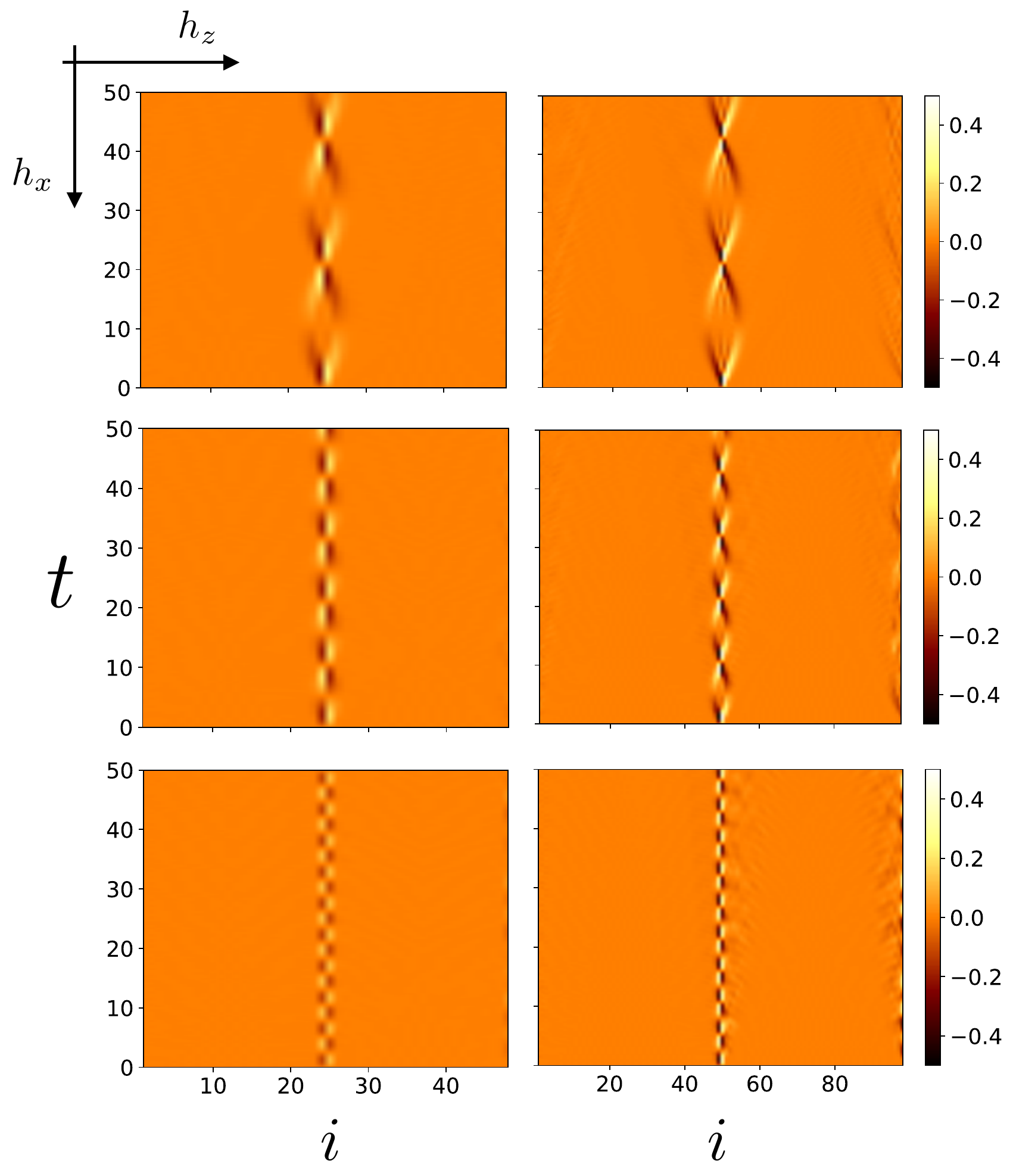}
\caption{Evolution of the energy density $ \langle \mathcal{H}_i(t) \rangle$ (left panel) and of the energy current density  $\langle \mathcal{J}_i(t) \rangle$ (right panel) profiles, governed by the Hamiltonian \eqref{Hamiltonian} starting from the inhomogeneous domain-wall state \eqref{initial_state}, obtained from TEBD simulations, for a range of increasing field values $h_z=0.2$ ($L=50$), $0.4$ ($L=100$) and $h_x=0.15,0.3,0.6$, varying as indicated by the axes. (Units are fixed such that $J=1$.) The same qualitative behavior as that illustrated here persists up to long times $t=10^3$. Note the oscillations of the profiles around the junction, with spatial amplitude $\propto h_z/h_x$ and frequency $\propto h_x$, while there is no evidence for the activation of transport.
}
\label{ShortEn}
\end{figure*}

\paragraph{Model and protocol.---}%
We consider a ferromagnetic quantum Ising chain with a transverse and a longitudinal magnetic field, $h_z$ and $h_x$, respectively: 
\be
H(h_z,h_x)=-J\sum_{i=1}^{L-1}\sigma^x_i\sigma^x_{i+1}-h_z\sum_{i=1}^L\sigma^z_i-h_x\sum_{i=1}^L\sigma_i^x.
\label{Hamiltonian}
\ee
Here $\sigma^{x,y,z}_i$ are the Pauli matrices acting on the site $i$, $J>0$ is the Ising exchange parameter, $L$ the (even) system size, and we consider open boundary conditions. 
For $h_x=0$, the model is exactly solvable in terms of free fermions \cite{refereeB.2,refereeB.1,schultzmattislieb} which, in the ferromagnetic phase with $\lvert h_z \rvert<J$, physically correspond to freely moving domain-walls (or kinks) connecting the two oppositely magnetized ground states with $\Braket{\sigma^x_j} \ne 0$. 
A finite $h_x\neq 0$ causes a non-perturbative modification of the spectrum of the elementary excitations: it selects as a ground state the one with $\Braket{\sigma^x_j}$ along $h_x$ and raises the energy of configurations with domains of reversed spins by an amount proportional to their extension. This corresponds to
a linear, V-shaped interaction potential between two consecutive kinks delimiting a domain, which therefore become \emph{confined} into composite objects called \emph{mesons}, in analogy with the low-energy limit of quantum chromodynamics. 
This modification of the spectrum has been studied both in the vicinity of the critical point $h_z \rightarrow 1$ exploiting field-theoretical methods 
\cite{McCoy, Muss, Zam}, 
and far away from it in the regime of low-density excitations for small $h_x$ \cite{Rut0}. 

In order to investigate transport processes it is convenient to consider a so-called \emph{inhomogeneous quench} \cite{inhomogeneous1,inhomogeneous2} in which two complementary subsystems are initially prepared in two different equilibrium states and then they are joined at time $t=0$, such that they evolve according to a common, homogeneous Hamiltonian. 
Here, we consider
%
a domain-wall initial state with  
a single kink in the middle of the chain 
which reads, in terms of the eigenstates $\upa_j$ and $\dna_j$ of $\sigma^x_j$, 
\be
|\Psi_0\rangle=\bigotimes_{j=1}^{L/2}\upa_j\bigotimes_{j=L/2+1}^L\!\!\dna_j \equiv \Ket{\uparrow\uparrow\dots\uparrow\uparrow\downarrow\downarrow\dots\downarrow\downarrow},
\label{initial_state}
\ee
and which is also an eigenstate of 
$H(0,h_x)$. 
At time $t>0$, the transverse field $h_z\ne0$ is suddenly switched on and we study 
the non-equilibrium evolution of the energy density profile $ \langle \mathcal{H}_j(t) \rangle$ as a function of $j$, where
\be
\mathcal{H}_j=-J \sigma^x_j\sigma^x_{j+1}-\frac{h_z}{2}\left(\sigma^z_j+\sigma^z_{j+1}\right)-\frac{h_x}{2}\left(\sigma^x_j+\sigma^x_{j+1}\right).
\label{endensity}
\ee
For $h_x=0$, the initial energy density $ \langle \mathcal{H}_j(t=0) \rangle$  is equal on the two sides of the junction, due to the $\mathbb{Z}_2$ symmetry. 
However, in the presence of a non-vanishing $h_x >0$, the chain acquires an initial macroscopic energy imbalance between the left (``cold'') part and the right (``hot'') part.  In particular, the latter may be viewed as  a ``false vacuum'' whose energy lies in the middle of the many-body spectrum, and may thereby be expected to decay into a finite density of traveling excitations upon activating the transverse field $h_z\ne0$, leading to a meltdown of the initial imbalance after a transient \cite{Rut1}. In the following we provide compelling evidence against 
this expectation.

\emph{Numerical analysis.---}In order to explore numerically the non-equilibrium evolution of the chain, we employ time-evolving block decimation (TEBD) simulations \cite{TEBD}.
It turns out that the entanglement grows slowly up to moderate values of the 
field $h_z \lesssim 0.4J$, 
which allows us to extend the simulations to long times $t_{\rm M}=10^3 J^{-1}$ with modest computational efforts, as in the case of Ref. \cite{Calabrese}. 
We investigate the behavior of  $\langle \mathcal{H}_j(t) \rangle$ [see Eq.~\eqref{endensity}]
and of the associated current $\langle \mathcal{J}_j(t) \rangle$, with
 \be
 \mathcal{J}_j= J h_z \left(  \sigma^{x}_{j-1}\sigma^{y}_{j} - \sigma^{y}_{j} \sigma^{x}_{j+1} \right),
 \label{curdensity}
\ee
for various values of $h_{x,z}$.  
The results of the simulations are illustrated in Fig.~\ref{ShortEn} only up to times 
$t = 50 J^{-1}$, as no qualitative differences are observed up to $t_{\rm M}$. 
In both the ``strong'' ($h_x\gg h_z$) and ``weak''  ($h_x \lesssim h_z$) confinement regime, 
energy transfer between the two halves of the chain is suppressed even at late times. 
As shown in Fig.~\ref{ShortEn}, the main dynamical effect of switching on $h_z$
is given by pronounced oscillations of the profiles around the position $j=L/2$ of the junction, with  characteristic emergent amplitudes and frequencies which depend on the values of the fields.
In particular, the energy current density is zero everywhere apart around the junction, where it oscillates between positive values (aligned with the energy gradient) and negative values (against the energy gradient).      
We emphasize that, within our protocol, an increase in the energy gradient between the two halves, caused by a stronger $h_x$, does not result in the activation of transport: on the contrary, it turns out that the oscillations at the junction acquire an even smaller amplitude (see Fig.~(\ref{ShortEn}) from top to bottom) . 
%

\begin{figure*}[!t]
\includegraphics[width=0.31\textwidth]{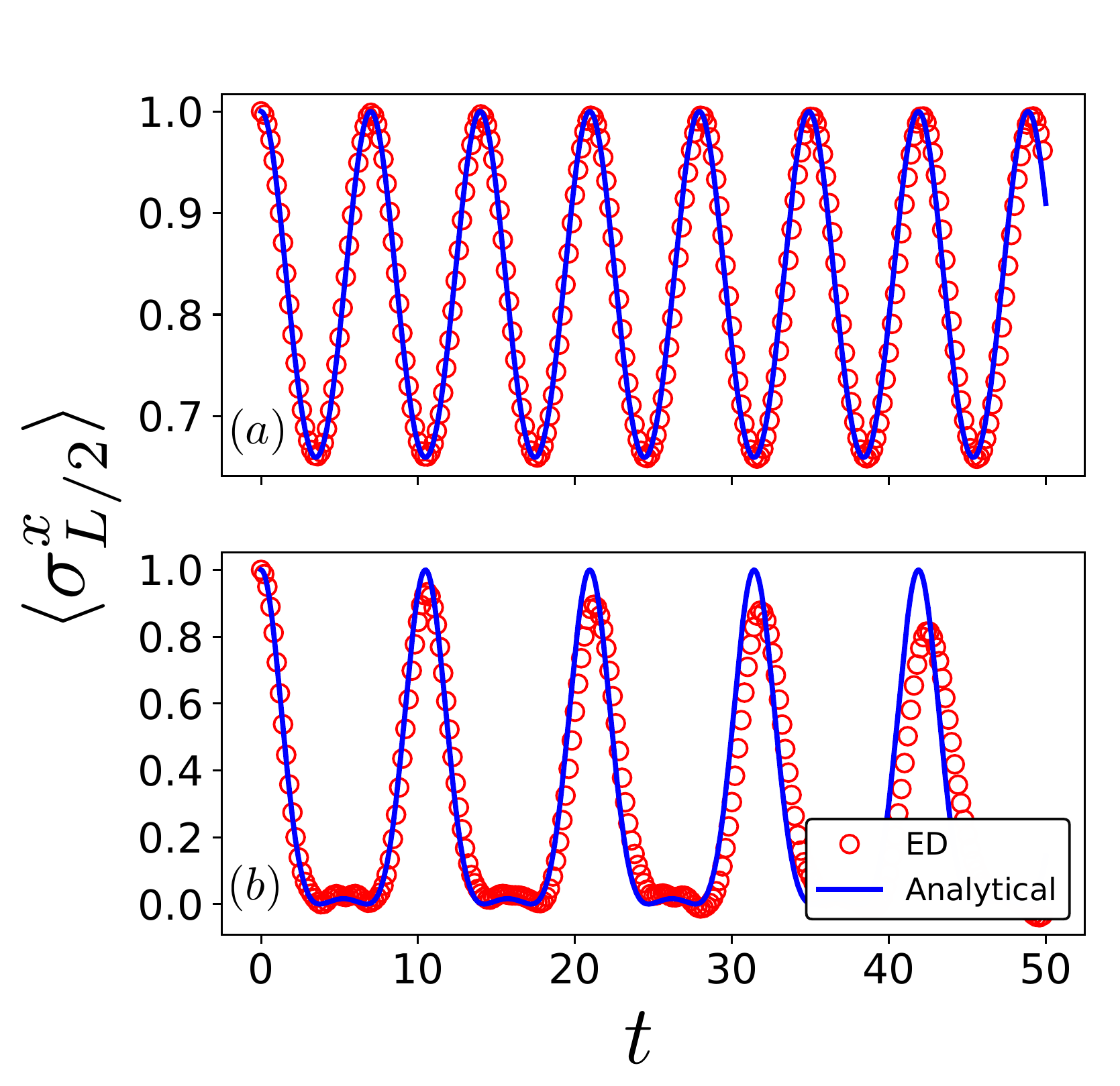}\hspace{0.2pc}
\includegraphics[width=0.31\textwidth]{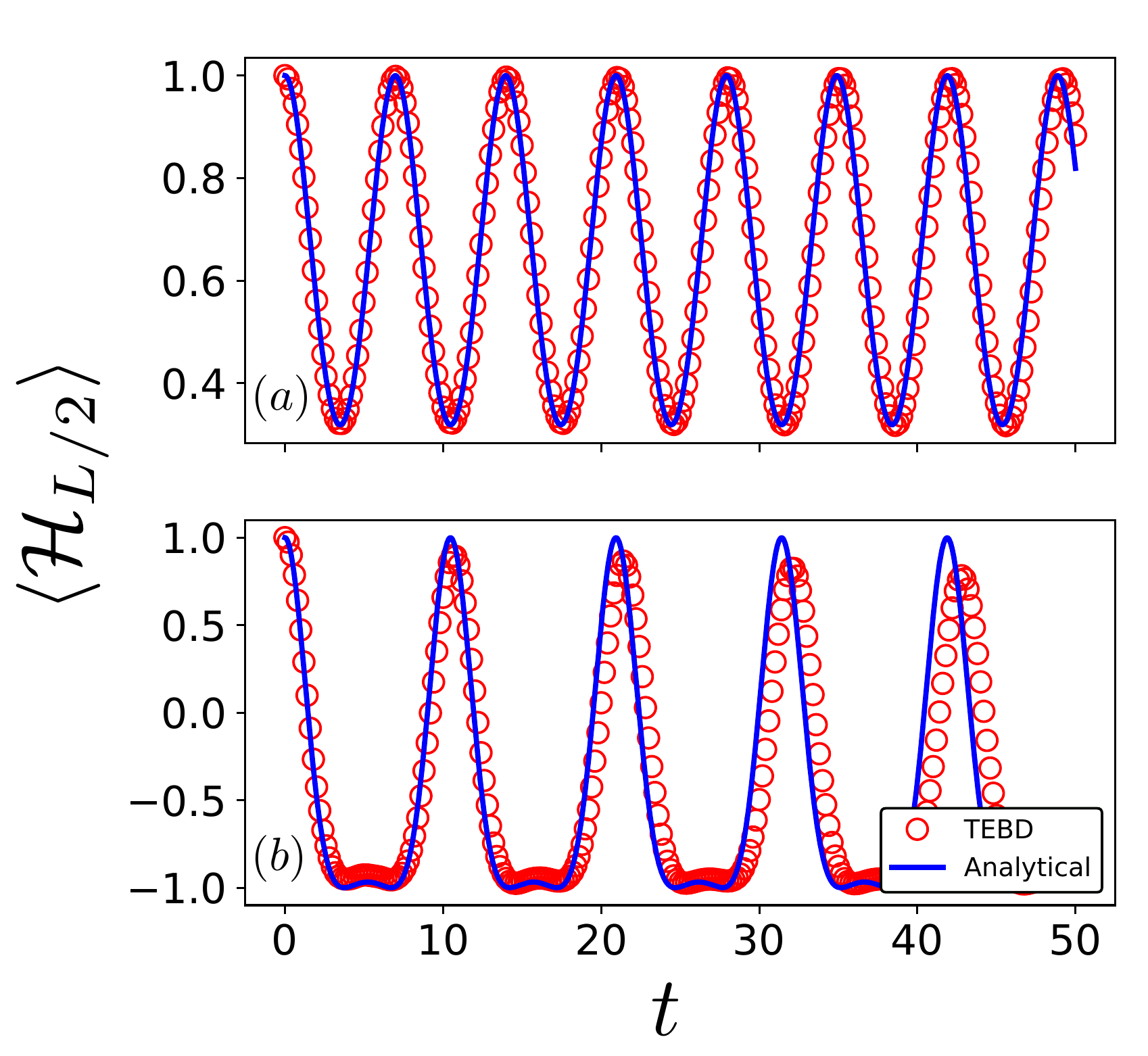}\hspace{0.2pc}
\includegraphics[width=0.31\textwidth]{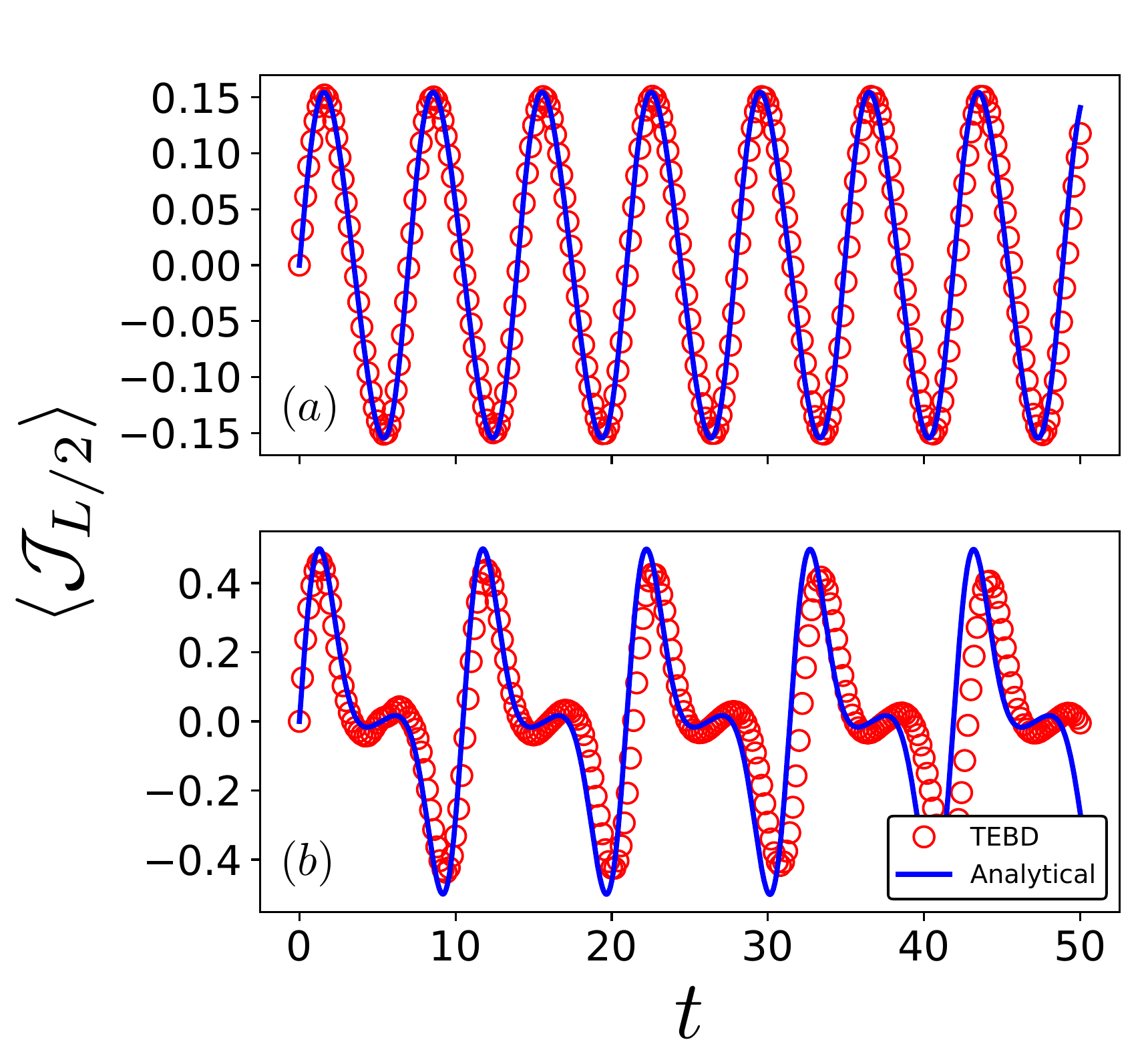}
\caption{Comparison between the numerical results $\langle \sigma^x_{L/2}(t) \rangle$, $\langle \mathcal{H}_{L/2}(t) \rangle$, $\langle \mathcal{J}_{L/2}(t) \rangle$ (symbols), and the analytical predictions
$m_{L/2}(t)$,
$e_{L/2}(t)$, $j_{L/2}(t)$ (solid lines)
 for the magnetization (left panel), energy density (central panel) and energy density current (right panel) respectively, at the junction $j=L/2$, as obtained from ED (with $L=16$) or TEBD (with $L=50$ or $100$), and from the effective single-particle model, respectively. These curves refer to $h_x=0.45$, $h_z=0.2$ (top row), and $h_x=0.3$, $h_z=0.4$ (bottom row). (Units are fixed such that $J=1$.) Note that discrepancies between symbols and solid lines appear as time increases, due to the neglected multi-kink processes. The associated time scale, however, increases upon decreasing $h_z$.}
\label{energy}
\end{figure*}

\emph{Effective dynamics.---}The oscillations of the profiles shown in Fig. \ref{ShortEn} may be interpreted as the quantum motion of the isolated kink initially localized at the junction, triggered by the transverse field $h_z\ne 0$. In fact, the kinetic energy associated with this motion   has a finite bandwith $ \sim h_z$ on the lattice, and therefore, because of energy conservation, the kink quasi-particle can travel,  in the linear confining potential $V(l)\sim - h_x l $, at most a distance $l_{\text{conf}} \sim h_z/h_x$ (confinement length scale), 
before bouncing back and oscillating. This phenomenon is analogous to the Wannier-Stark localization of electrons in a one-dimensional crystal subject to a constant electric field \cite{wannier}. 

In order to rationalize the above intuition and make a quantitative treatment of the evolution of the profiles, we propose a simple analytical approach based on dressing the meson quasi-particles perturbatively in the transverse field $h_z \ll J$, with arbitrary $h_x$. 
(This regime differs from the one $h_x \ll J$, $h_z < J$, of validity of the semiclassical technique of Ref. \cite{Rut0}.) The approximation consists in neglecting the creation of new quasi-particles, which, in our setup, only affect the quantum fluctuations in the two homogeneous bulks away from the junction, as recognized in Refs. \cite{Calabrese,Rut0}. In fact, we show that the dynamics at the junction is very well captured within this scheme up to moderate values of $h_z$.

In the spirit of an effective quasi-particle description of mesons \cite{cinesi}, we map the motion of the isolated kink onto the problem of a \emph{single quantum particle} hopping on a one-dimensional lattice, by projecting the many-body Hilbert space onto the single-kink linear subspace \cite{SupplMat}. As the numerical results indicate (see Fig. \ref{ShortEn}), for $h_z \ll J$ and arbitrary $h_x$ \footnote{
Note, however, that resonances occur at particular values of $h_x$, commensurate with $2J$. Correspondingly, it costs no energy to break a single meson into multiple mesons by flipping individual spins. These transitions cause quantitative but not qualitative modifications to the evolution of the energy profile, which are not captured by the single-kink subspace projection discussed here. Evidence of this aspect can be found in the Supplemental Material.}
the dynamics can be approximated within this subspace, spanned by the states $\{\Ket{n}\}$ with a single domain-wall located between sites $n$ and $n+1$, with $n=1,2,\dots,L-1$. The corresponding unperturbed energy eigenvalues are $E_n=2J+2h_x (L-n)+E_{\text{GS}}$, where $E_{\text{GS}}=-J(L-1)-h_x L$ is the ground energy of the chain. The resulting matrix elements $\Braket{n|H(h_z,h_x)|m}$ of the Hamiltonian  (\ref{Hamiltonian})  read $E_{\text{GS}}\delta_{n,m} + \left(H^{\textnormal{eff}}\right)_{nm}$, with 
\be
\left(H^{\textnormal{eff}}\right)_{nm}=\left[2J+2(L-n)h_x\right]\delta_{n,m}-h_z(\delta_{n,m+1}+\delta_{n,m-1}).
\label{effectiveham}
\ee
We note that the off-diagonal perturbation produces an effective hopping amplitude for the kink quasi-particle. Accordingly, the effective Hamiltonian  $H^{\textnormal{eff}}$ describes the dynamics in terms  of a single particle hopping in a one-dimensional lattice in the presence of a linear potential, where the state of the particle is described by a vector $\{\psi_n\}$ with $n=1,2,\dots,L-1$. The absolute value squared of  the $n$-th component of the wavefunction $\psi_n(t)$ is equal to the probability that the particle is at site $n$ at time $t$. Within this picture, the initial state in Eq. \eqref{initial_state} maps to $\psi_n(0) = \delta_{n,L/2}$, corresponding to a particle completely localized at the junction between the two chains. 
Similarly, the magnetization $\langle \sigma^x_j(t) \rangle$ at site $j$ and time $t$ can be expressed \cite{SupplMat} within this single-particle picture as
\be
m_j(t) \equiv 1-2\sum_{n=1}^{j-1} \big\lvert \psi_n (t)\big\rvert^2,
\ee
where $\psi_n (t) = \sum_m (\mbox{exp}(-iH^{\textnormal{eff}}t))_{nm} \psi_m(0)$ is the time evolved state within the projected space.
 
In order to test the accuracy of our approximation, we compare the dynamics obtained from the above effective single-particle problem with the exact dynamics generated by $H$ [see Eq. \eqref{Hamiltonian}] in the full many-body Hilbert space, starting from the domain-wall initial state $\Ket{\Psi_0}$ of Eq. \eqref{initial_state} as obtained via both exact diagonalization (ED) and TEBD techniques \footnote{In this case, the simulations based on exact diagonalization of the Hamiltonian can be pushed until unexpected long times because finite-size effects such as revivals are suppressed, due to the fact that excitations are confined \cite{Calabrese}.}. The comparison between $m_{L/2}(t)$ and $\langle \sigma^x_{L/2}(t) \rangle$  is shown in Fig. \ref{energy}.  In particular, we observe that the agreement is fairly good up to moderate values of the transverse field $h_z \lesssim 0.4 J$.

Similarly, the relevant non-equilibrium profiles of the energy and energy current densities can be studied within the above effective single-particle description. This is achieved by projecting the energy density $\mathcal{H}_j$ at site $j$ in Eq. \eqref{endensity} onto the single-kink subspace,
\begin{multline}
 \left(\mathcal{H}_j^{\textnormal{eff}}\right)_{nm}=\frac{1}{2}\left[J(2 \delta_{j,n} -1)-h_x \mbox{sgn}(n-j)  \right]\delta_{n,m} \\-\frac{h_z}{2}\left(\delta_{j,m+1}+\delta_{j+1,m+1} \right)\delta_{n,m+1} + (m\leftrightarrow n),
 \end{multline} 
where the sign function $\mbox{sgn}(x)$ equals $1$ for $x>0$, $-1$ for $x<0$ and $0$ for $x=0$. From the continuity equation
\be
\frac{d\mathcal{H}^\textnormal{eff}_j}{dt}=i[H^{\textnormal{eff}},\mathcal{H}^\textnormal{eff}_j]= \mathcal{J}^{\textnormal{eff}}_j- \mathcal{J}^{\textnormal{eff}}_{j+1},
\ee
we can infer the corresponding effective expression for the energy current density operator $\mathcal{J}_j$ at site $j$, i.e.,
\be
\begin{split}
\left(\mathcal{J}^{\textnormal{eff}}_j\right)_{nm}  =& \; 2i J h_z\delta_{n,m+1} \; \delta_{m,j-1}
\\ & -\frac{i}{2}h_z^2  \delta_{m,j-2}\;   \delta_{n,m+2}  -\frac{i}{2}h_z^2  \delta_{m,j-1}\;   \delta_{n,m+2}  \\ & - (m\leftrightarrow n)
\end{split}
\ee
The time-dependent expectation value of the energy density at site $j$ within this single-particle picture can therefore be written  as
\be
e_j(t) \equiv \sum_{n,m}\psi_n^{\ast}(t)\left(\mathcal{H}_j^\textnormal{eff}\right)_{nm}\psi_m(t)
,
\ee
with an analogous expression for the current $j_j(t)$, in terms of $\mathcal{J}^{\textnormal{eff}}_j$.
In Fig. \ref{energy} we compare the time evolution of $e_{L/2}(t)$ and $j_{L/2}(t)$ with the corresponding exact quantities $\langle \mathcal{H}_{L/2}(t) \rangle$ and $\langle \mathcal{J}_{L/2}(t) \rangle$ as obtained from the TEBD simulations. (One can show that the spectrum of the effective Hamiltonian \eqref{effectiveham} consists of multiples of $2h_x$, which results in exactly periodic behavior of the blue lines in Fig. \ref{energy}). It is remarkable that, in spite of the simplicity of this approach, the agreement is excellent for small values $h_z = 0.2 J$ of the transverse field, whereas for larger values $h_z = 0.4 J$, small quantitative discrepancies appear, still retaining a fairly good qualitative agreement.

\emph{Conclusions.---}%
In a homogeneous quench, the confinement  of excitations has been recently shown to hinder the spreading of correlations in  the quantum Ising chain  \eqref{Hamiltonian} with both transverse and longitudinal magnetic fields \cite{Calabrese}. (Anomalous non-equilibrium evolution had already been reported in the same model, but within a different regime of parameters, in Refs. \cite{cirac} and \cite{cinesi}.) In this work, we have shown that this phenomenon has significant consequences even in inhomogeneous setups, as it can lead to suppression of energy transport. 
 This lack of transport in the presence of an initial gradient actually mirrors the fact that the spatial inhomogeneity in the longitudinal magnetization persists at long times, meaning that the system fails to locally relax to the thermal ensemble up to the largest accessible and explored times, $t_M= 10^3 J^{-1}$, which are longer than those currently accessible in experiments.

We emphasize that in the problem discussed here the specific choice of the class of inhomogeneous initial states plays an important role. As we have shown, the non-equilibrium dynamics are accurately captured by the boundary Bloch oscillations  of a single macroscopically large ``meson''. Based on extensive numerical work, it has been recently suggested in Ref. \cite{Konik},  that the Hamiltonian \eqref{Hamiltonian} is characterized by a pattern of atypical energy eigenstates with non-thermal features carrying over to the thermodynamic limit, which violate the eigenstate thermalization hypothesis \cite{ETH}. In this light, our results may represent a dynamical manifestation of this phenomenon. In particular, if the initial states have significant overlap with those ``single-meson'' non-thermal eigenstates, the initial inhomogeneity would persist to \emph{infinite} time. However, we argue that more general initial states, with magnetic domains separated by distances much larger than the confinement length scale, would also retain their inhomogeneity for a correspondingly long time. 
The phenomenon reported here may be interpreted as a dramatic slowdown or suppression of ``string breaking'' in one-dimensional quantum models with confinement of excitations. As such, we expect it to occur rather generically in this context, e.g., in XXZ spin chains \cite{SchollwockRealTimeConfinement,WangConfinement,EsslerConfinement,Rut2}, one-dimensional extended Bose-Hubbard models \cite{GiappiHiggs,GiappiGlassy}, spin-$1$ quantum chains \cite{S1Confinement1,S1Confinement2} and systems with long-range interactions \cite{GorshkovConfinement,BojanDW}, as well as lattice models of quantum electrodynamics \cite{SuraceRydberg}. 
Similarly, we observe that recent works have reported the occurrence of localization phenomena --- and thereby of suppression of information spreading --- within the context of lattice gauge theories, where confinement of elementary excitations naturally arises as well \cite{Montangero,CiracDemlerVariational,SuraceRydberg,NandkishoreMBLConfinement,Gauge-inv,Smith,GiappiGlassy}. 
In future work we plan to investigate the very origin of this seemingly ubiquitous phenomenon, as well as to address the important problem of estimating the relaxation time scales.

\emph{Acknowledgments.---}%
We acknowledge useful discussions with P. Calabrese, G. B. Mbeng, J. Moore, G. Pagano, N. Robinson, and S. B. Rutkevich.
We thank G. Giudici for technical help with the numerical codes. The work of M. C. was supported by the European Union's Horizon 2020 research and innovation programme 
under the Marie Sklodowska-Curie Grant Agreement No. 701221.



\begin{widetext}

\newpage 

\begin{center}
\textbf{\large Supplemental Material: \\ Suppression of transport in non-disordered quantum spin chains \\ due to real-time confinement}
\end{center}
\setcounter{equation}{0}
\setcounter{figure}{0}
\setcounter{table}{0}
\setcounter{page}{1}
\makeatletter
\renewcommand{\theequation}{S\arabic{equation}}
\renewcommand{\thefigure}{S\arabic{figure}}
\renewcommand{\bibnumfmt}[1]{[S#1]}
\renewcommand{\citenumfont}[1]{S#1}

This supplemental material is organized as follows: in Sec. I we motivate the effective single-particle approach used in the main text, in Sec. II we define the restricted single-kink subspace and derive the effective Hamiltonian which governs the time evolution within this subspace, in Sec. III we study the corresponding dynamics of the energy and energy current densities, while in Sec. IV we briefly discuss the range of validity of this approach and its limitations.

\section{Motivation for the effective single-particle approach}
\label{firstsection}

As briefly outlined in the main text, 
we focus on the dynamics generated by $H(h_z,h_x)$ in Eq. (1), within a perturbative expansion for $h_z \ll J$ and
 therefore we first discuss the structure of the spectrum of the unperturbed Hamiltonian $H(h_z=0,h_x)$. 
Its eigenstates can be written 
in terms of the eigenvectors $|\uparrow\rangle = \frac{1}{\sqrt{2}}(1,1)$, $|\downarrow\rangle = \frac{1}{\sqrt{2}}(1,-1)$ of the Pauli matrix $\sigma^x$. Examples of these eigenstates are reported in  Fig. \ref{stati}, where the various arrows correspond to the lattice sites of the chain.
\begin{figure}[h]
\centering
\includegraphics[scale=0.6]{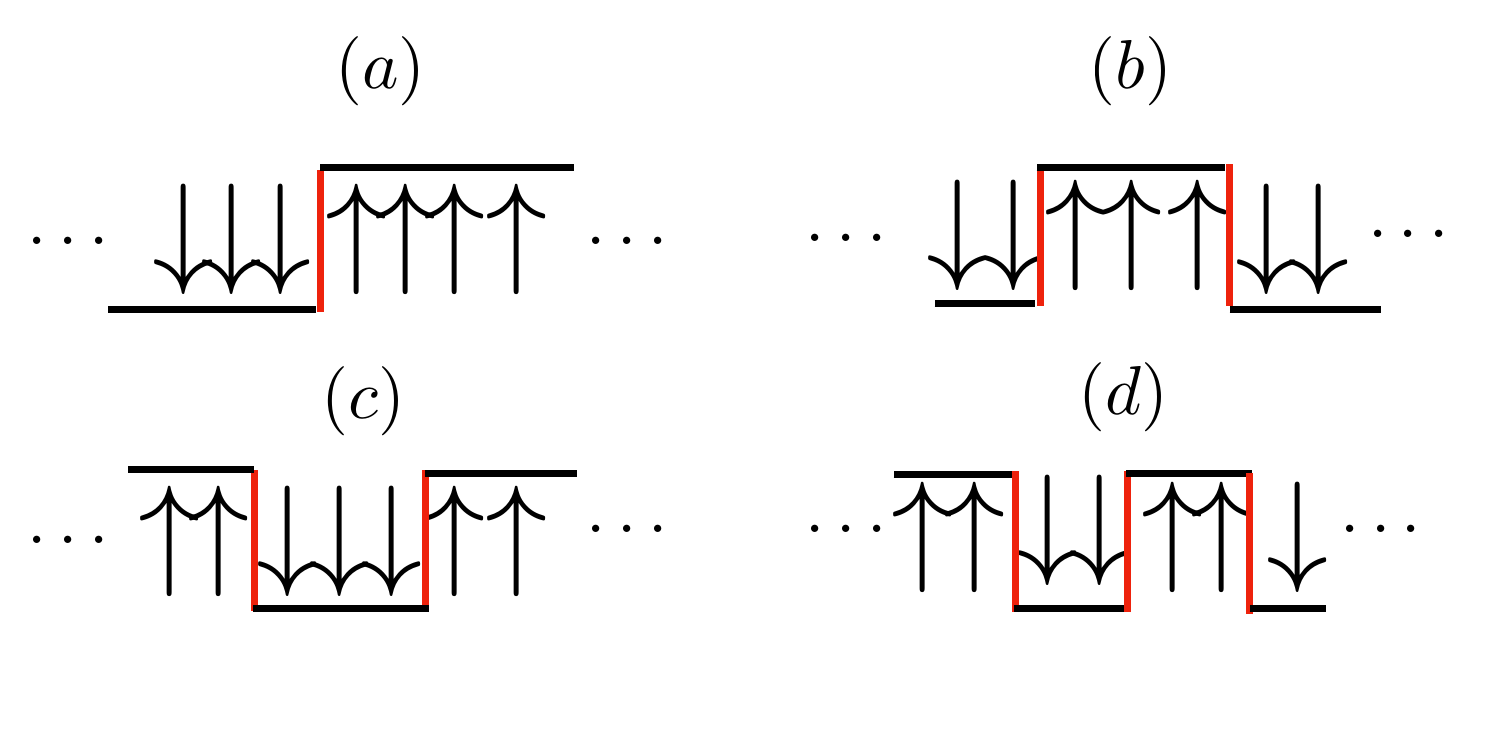}
\caption{
Schematic representation of some excited states of $H(h_z=0,h_x)$, where the arrows indicate the eigenvectors $|\uparrow\rangle$, $|\downarrow\rangle$ of $\sigma^x$ at the various lattice sites of the chain while the red vertical bars indicate the occurrence of domain-walls or kinks.
} 
\label{stati}
\end{figure}
 In particular, the ground state $\Ket{GS}$ of $H(0,h_x)$ is
\be
\Ket{GS}=\Ket{\uparrow\uparrow\dots\uparrow\uparrow}
\ee
for $h_x>0$ (a similar analysis can be done for $h_x<0$).
The energy levels $E(k,l)$ of the excited states can be characterized by two quantum numbers, namely the total number $k$ of kinks (or domain-walls) and the total number $l$ of reversed spins (i.e., arrows pointing downward), as 
\be
\label{eq:spectrum}
E(k, l)=E_{\text{GS}}+2J k+2h_x l,
\ee 
with $E_{\text{GS}} = -J (L-1) - h_x L$.
Note that the corresponding eigenspaces are highly degenerate, because energy is unchanged upon separately shifting each  domain with consecutively reversed spins (``meson'')  by arbitrary distances, retaining the same number of kinks $k$.
Given this structure, we can pictorially arrange the energy levels of the excited states in bands labelled, e.g., by the number $k$ of kinks, as reported in Fig. \ref{sketch}. 
\begin{figure}[t]
\centering
\includegraphics[scale=0.5]{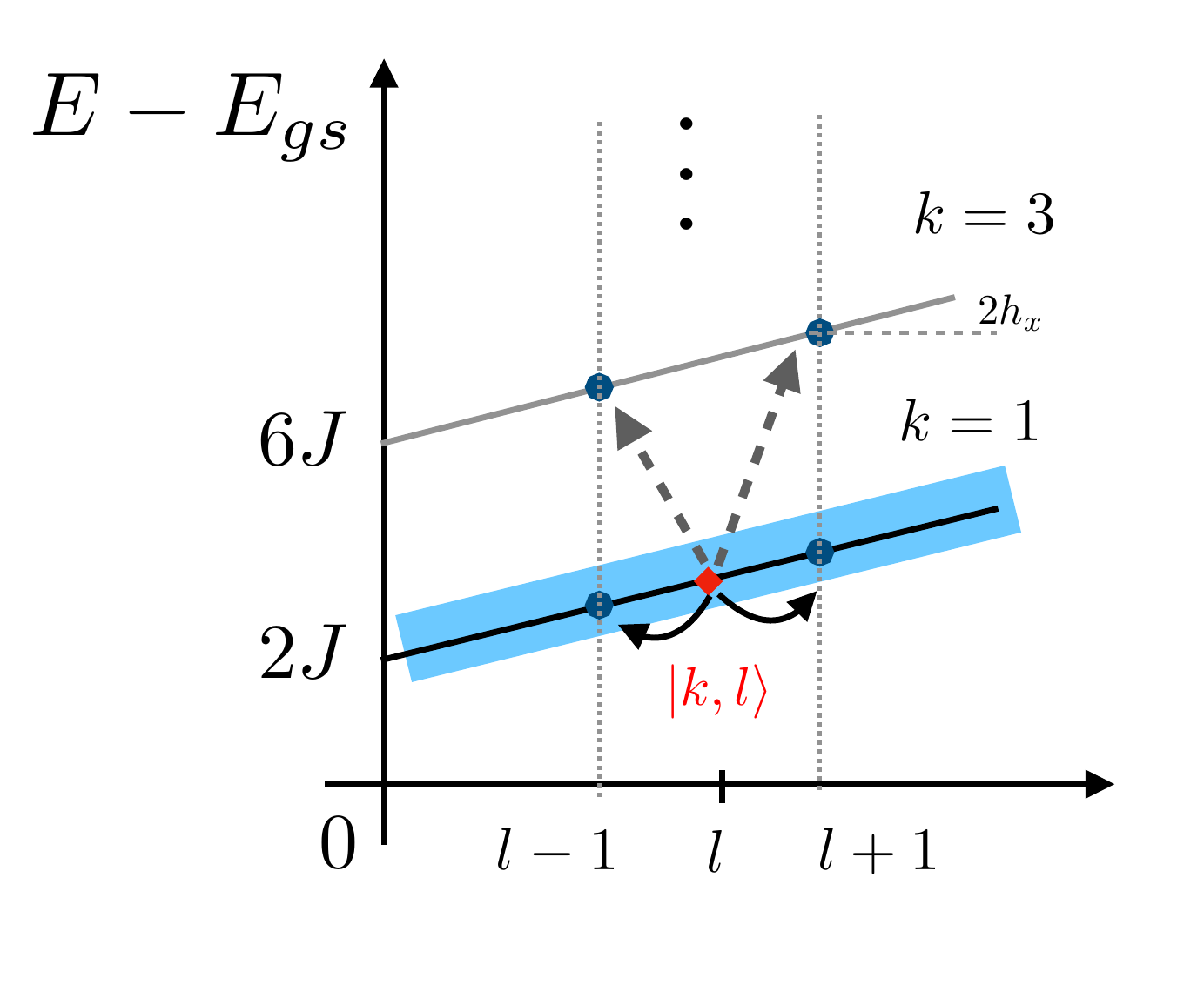}
\caption{Energy spectrum $E-E_{\text{GS}}$ of the excited states of $H(h_z=0,h_x)$. 
From the initial state $\Ket{k,l}$ marked in red, belonging to the single-kink ($k=1$) band shaded in blue, 
the  allowed transitions at first order in $h_z$ are indicated by the various arrows and they correspond to single spin flips.
If $h_x \ll J$, the transitions (dashed arrows) occurring outside the single-kink band  are suppressed with respect to those (solid arrows) occurring within it.
}
\label{sketch}
\end{figure}

When the perturbation
\be
\label{eq:perturbation}
V 
=-h_z\sum_{i=1}^L\sigma^z_i
\ee
is switched on, transitions between states  $\Ket{k,l}$ and $ \Ket{k', l'}$ become possible, where $\Ket{k,l}$ ($\Ket{k',l'}$) denotes a representative state in the eigenspace identified by $k$ and $l$ ($k'$ and $l'$).
For the single-kink ($k=1$) initial states considered in the main text, the possible transitions occurring up to the first order in $h_z$ are schematically shown in Fig. \ref{sketch}. In particular, they change $l$ by one, while $k$ can either remain constant or increase by two
(in this discussion, for simplicity we disregard spin flips occurring at the boundary of the chain), i.e.,
\begin{enumerate}
\item $|k, l\rangle\to |k, l\pm1\rangle$;
\item $|k,l\rangle \to |k+2, l\pm1\rangle$.
\end{enumerate}
At the first order in perturbation theory, the associated long-time transition amplitudes   are proportional to the matrix element $\Braket{k',l'|V|k,l}$ of the perturbation  between the two states divided by their energy difference $E(k',l')-E(k,l)$ (see, e.g., Ref. \cite{sakurai}), i.e., to
 \begin{enumerate}
\item $h_z/ ( \pm 2 h_x)$;
\item $h_z/(4J \pm 2 h_x)$.
\end{enumerate}
For $h_x \ll J$, the probability amplitude of process 1 above is much larger than that of process 2. Accordingly, transitions between single-kink states dominate the dynamics, which therefore can be conveniently projected onto the subspace spanned by these states (see Fig. \ref{sketch}). 
If the value of $h_x$ is, instead, comparable to or larger than $J$, the transitions discussed above are generically suppressed as long as the corresponding energy denominators do not vanish. For further details on the range of validity and on the limitations of the present single-particle approach, see Sec. IV.

\section{Effective dynamics within the single-kink subspace}

Within the perturbative regime $h_z \ll J$, the dynamics of the system prepared in the state $\Ket{\Psi_0}$ involve only single-kink states. Accordingly, we can project it onto the $L-1$-dimensional subspace spanned by the states $\{ |n\rangle \}$  with a single kink located between the lattice sites $n$ and $n+1$, i.e.,
\begin{equation}
| n \rangle = \bigotimes_{i=1}^{n}|\uparrow\rangle_i \bigotimes_{i=n+1}^L|\downarrow\rangle_i \quad \mbox{with} \quad n=1,2,\dots,L-1. \label{basis}
\end{equation}
As already stated in Sec. \ref{firstsection}, for simplicity we assume $h_x>0$, such that the single ``meson'' in the state $| n \rangle$ is given by a block of $L-n$  spins reversed with respect to the longitudinal field direction $x$. Note that the initial state $| \Psi_0 \rangle$ in Eq. (2) of the main text corresponds to $|L/2 \rangle$. 
These states are eigenstates of the Hamiltonian $H(0,h_x)$,
\begin{equation}
H(0,h_x)|n \rangle = [2J +2(L-n)h_x + E_{\text{GS}}] |n \rangle,
\end{equation} 
where  the ground state energy $E_{\text{GS}} = -J(L-1)-h_x L$ of $H(0,h_x)$ is an additive constant 
which we neglect in what follows. 
The transverse field $h_z$ provides a kinetic energy to the kinks and, as a matter of fact, it allows transitions from the state $|n\rangle$ to states $|n \pm 1 \rangle$: since $\sigma^z_j |\uparrow \rangle_j = |\downarrow \rangle_j$ (and $\sigma^z_j |\downarrow \rangle_j = |\uparrow \rangle_j$), one has
\begin{equation}
(V)_{nm} = -h_z \delta_{n,m+1} -h_z \delta_{n,m-1},
\end{equation}
where $V$ is defined in Eq. \eqref{eq:perturbation}.
Accordingly,  the effective Hamiltonian $H^{\text{eff}}$ governing the evolution of the system in the single-kink subspace reads
\begin{equation}
 \left(H^{\textnormal{eff}}\right)_{nm} 
 = \left[2J+2(L-n)h_x\right]\delta_{n,m}-h_z(\delta_{n,m+1}+\delta_{n,m-1}),
\end{equation}  
which is Eq. (5) in the main text. This Hamiltonian determines the time evolution of observables in the single-kink subspace and it can therefore be  used to study the dynamics of the magnetization $\sigma^x_j$ at each lattice site $j$. Within the single-kink subspace spanned by the states in Eq. \eqref{basis}, a generic vector $|\Psi \rangle$ can be written as 
\begin{equation}
|\Psi \rangle = \sum_{n=1}^{L-1} \psi_n | n \rangle,
\end{equation}
with suitable coefficients $\psi_n$.
The average longitudinal spin component $\langle \sigma^x_j \rangle$ at site $j$ can therefore be written as
\begin{equation}
\langle \sigma^x_j \rangle = \langle \Psi| \sigma^x_j |\Psi \rangle \equiv m_j = 1- 2 \sum_{n=1}^{j-1} |\psi_n|^2,
\end{equation}
where we have exploited the identity 
\begin{equation}
\sigma^x_{j} |n \rangle = u(n-j) |n \rangle \label{sigmax},
\end{equation}
in which the unit step $u(x)$ is defined such that $u(x\ge0)=+1$ and $u(x<0)=-1$.
Hence, the dynamics of the magnetization $m_j$ is readily obtained as 
\begin{equation}
m_j(t) = 1- 2 \sum_{n=1}^{j-1} |\psi_n(t)|^2,
\end{equation}
in terms of the coefficients $\psi_n (t) = \sum_{m}(\mbox{exp}(-iH^{\textnormal{eff}}t))_{nm} \psi_m (0)$ of the time-evolved state. 
In the problem considered in the main text, the initial state is a kink localized in the middle of the chain, corresponding to $| L/2 \rangle$, and therefore   $\psi_n(0) = \delta_{n,L/2}$.


\section{Representation of transport observables in the single-kink subspace}

The same procedure as the one described in Sec. II can be repeated for the energy density, Eq. (3) in the main text. In particular, by using the fact that $\sigma^y_j |\uparrow\rangle_j =-i|\downarrow\rangle_j$ ($\sigma^y_j |\downarrow\rangle_j = i|\uparrow\rangle_j$), we obtain 
\begin{equation}
\left(\sigma^y_{j}\right)_{nm} = i \delta_{j,n+1}\delta_{n,m+1} -i \delta_{j,n}\delta_{n,m-1} \label{sigmay},
\end{equation}
which, together with Eq. \eqref{sigmax}, yields Eq. (7) of the main text. 
The  energy density operator $\mathcal{H}^{\text{eff}}_j$ at site $j$ obtained in this way is a well-defined density, since its sum over the lattice sites of the chain renders the projected Hamiltonian $H^{\text{eff}}$, i.e.,
\begin{equation}
H^{\text{eff}} = \sum_{j=1}^{L-1} \mathcal{H}^{\text{eff}}_j. 
\end{equation}  
The energy current density $\mathcal{J}^{\text{eff}}_j$ at site $j$  in the single-kink subspace can be conveniently defined by writing the commutator between $\mathcal{H}^{\text{eff}}_j$ and $H^{\text{eff}}$ as a divergence of a local current, as done in Eqs. (8) and (9) of the main text. 
Note that the energy current $\mathcal{J}^{\text{eff}}_j$ obtained in this way differs from the one which would have been obtained by  projecting directly the current operator in Eq. (4) of the main text on the single-kink subspace, with an analogous prescription as the one followed above for $\mathcal{H}^{\text{eff}}_j$. In fact, this procedure would have led to 
\begin{equation} 
(\widetilde{\mathcal{J}}_j)_{nm}  =  2i J h_z\delta_{n,m+1}\delta_{m,j-1}+\textnormal{h.c.}, \label{projectedcurrent}
\end{equation}
which differs from $\mathcal{J}_j^{\text{eff}}$ in Eq. (9) of the main text by terms of order $h_z^2$. This can be realized by considering the continuity equation which defines the energy density and current at the operator level, i.e.,
\begin{equation}
\frac{d \mathcal{H}_j}{d t} = i [H, \mathcal{H}_j] = \mathcal{J}_j - \mathcal{J}_{j+1},
\end{equation}
with $H$, $\mathcal{H}_j$, $\mathcal{J}_j$ given by Eqs. (1), (3) and (4), respectively, of the main text. By projecting this equation over the single-kink subspace spanned by the indices $n,m = 1,2,\dots,L-1$, one finds 
\begin{eqnarray}
(\widetilde{\mathcal{J}}_j)_{nm} - (\widetilde{\mathcal{J}}_{j+1})_{nm} =& i([H,\mathcal{H}_j])_{nm}  \nonumber \\
                                                          \neq & i [(H^{\text{eff}}),(\mathcal{H}^{\text{eff}}_j)]_{nm} \nonumber \\
                                                          \neq & (\mathcal{J}^{\text{eff}}_j)_{nm} - (\mathcal{J}^{\text{eff}}_{j+1})_{nm}.  
\end{eqnarray}  
On the other hand, in the regime of validity of our approximation scheme, corresponding to $h_z \ll J$, terms of order $h_z^2$ in Eq. (9) are negligible and there is actually no difference in using the definition of Eq. \eqref{projectedcurrent} or the one in Eq. (9) of the main text. The latter, however, has been preferred in order to have an effective energy current operator $\mathcal{J}_j^{\text{eff}}$ which satisfies the continuity equation (8) of the main text.

\section{Range of validity and limitations}

The agreement between the effective dynamics within the single-kink subspace discussed in the previous sections and the exact dynamics in the full many-body Hilbert space is fairly good for arbitrary values of $h_x/J$, as shown in Fig. (2) of the main text and for a wider range of parameters in Fig. \ref{fig:arbitraryhx}. 

\begin{figure}[t]
\centering
\includegraphics[scale=0.5]{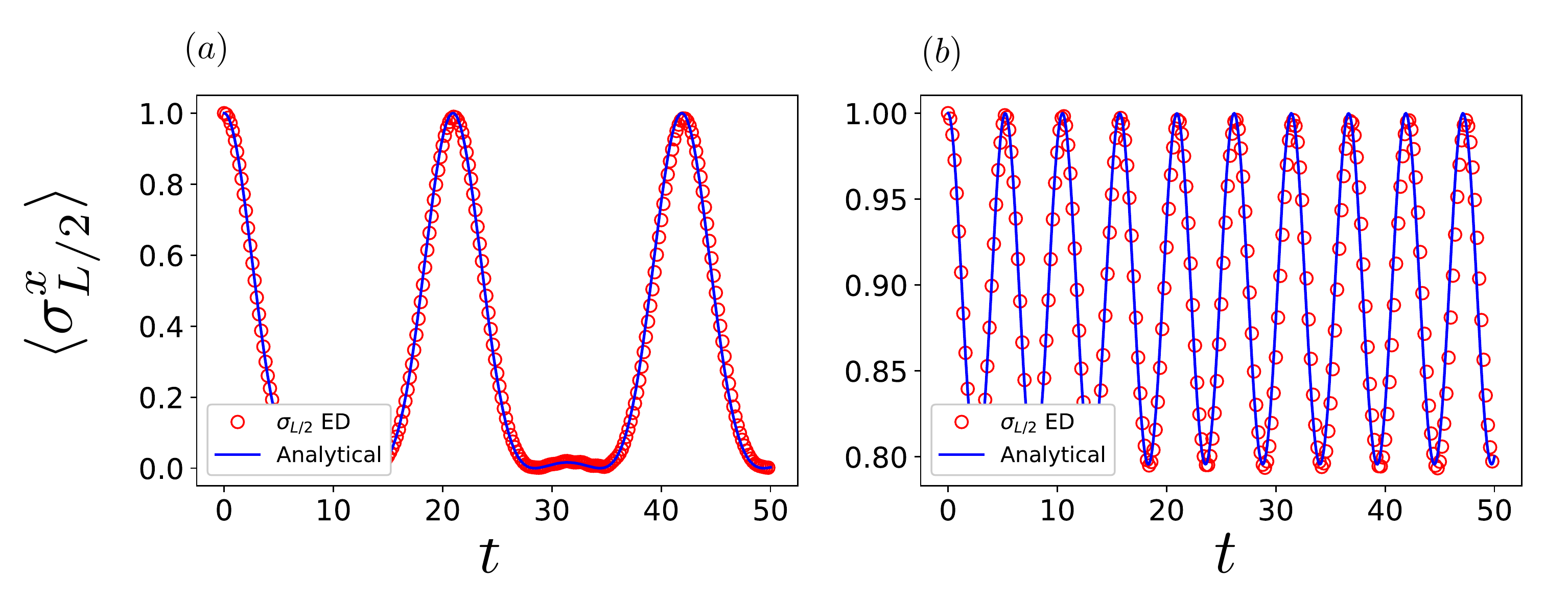}
\caption{
Comparison between the effective dynamics within the single-kink subspace (solid line) and the exact dynamics in the full many-body Hilbert space [symbols, determined via exact diagonalization (ED) with $L=16$] for both small ($h_x=0.15 $, left panel) and intermediate ($h_x=0.6 $, right panel) values of $h_x$. Here $h_z =0.2$ and units are fixed such that $J=1$. Although the data shown here refer to the average magnetization $\Braket{\sigma^x_{L/2}(t)}$ at the center of the chain, a similar agreement is observed for the average energy density $\Braket{\mathcal{H}_{j}(t)}$ and energy current density $\Braket{\mathcal{J}_{j}(t)}$ up to similar times.
}
\label{fig:arbitraryhx}
\end{figure}

However, 
if $h_x$ and $J$ are commensurable, degeneracies occur in the unperturbed energy spectrum in Eq. \eqref{eq:spectrum}. In particular, this happens when the energies $E(k,l)$ and $E(k',l')$ of two states $\Ket{k,l}$ and $
\Ket{k',l'}$, respectively, are equal, i.e., when 
\be
E(k,l) - E(k',l')=2J(k-k') +2h_x (l-l') =0
.
\ee 
If two such states are connected by $n$ spin flips, then resonances occur at the $n$-th order in perturbation theory, and therefore their effect becomes manifest only at a correspondingly long time scale. 

For instance, at the first order in $h_z$,  the process 2 above is resonant when $h_x \simeq 2J $. In this case, it costs no energy to break the single initial meson into multiple mesons via single spin flips. During the evolution, a finite density of isolated reversed spins is thus generated, thereby lowering the average local magnetization. 
Since these states have $k>1$, they no longer belong to the single-kink subspace and the effective single-particle description employed in the main text is therefore not expected to properly capture the resulting dynamics, 
as indeed demonstrated in Fig. \ref{wrongdyn}. 
However, as long as the perturbation $h_z$ is small, these reversed domains have a low spatial density  and therefore they all have zero momentum (cf. Ref. [63] of the main text). Accordingly, in this perturbative regime, transport is not expected to be activated by the presence of these resonances, as shown in Fig. \ref{fig:resonance}. 

\begin{figure}[t]
\centering
\includegraphics[scale=0.5]{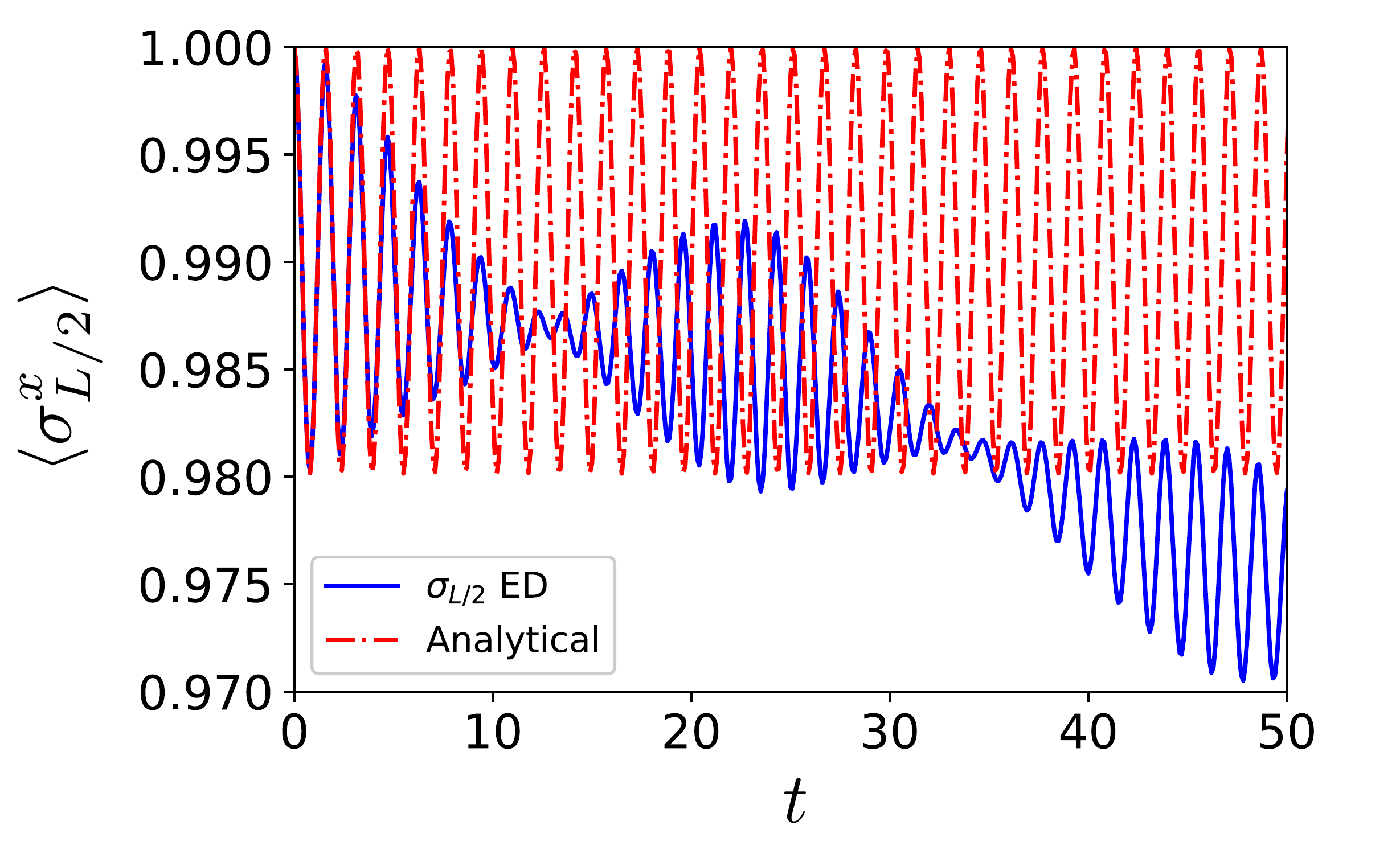}
\caption{
Comparison between the numerical data [determined by exact diagonalization (ED) with $L=16$] of the magnetization $\Braket{\sigma^x_{L/2}(t)}$ at the junction $j=L/2$ and the corresponding analytical prediction $m_{L/2}(t)$. 
At variance with what is observed in Fig. \ref{fig:arbitraryhx}, a qualitative discrepancy emerge between the two curves, due to resonances at first order in perturbation theory. These curves refer to $h_x=2$ and $h_z=0.2$, where the units are fixed such that $J=1$.
}
\label{wrongdyn}
\end{figure}

\begin{figure}[t]
\centering
\includegraphics[scale=0.45]{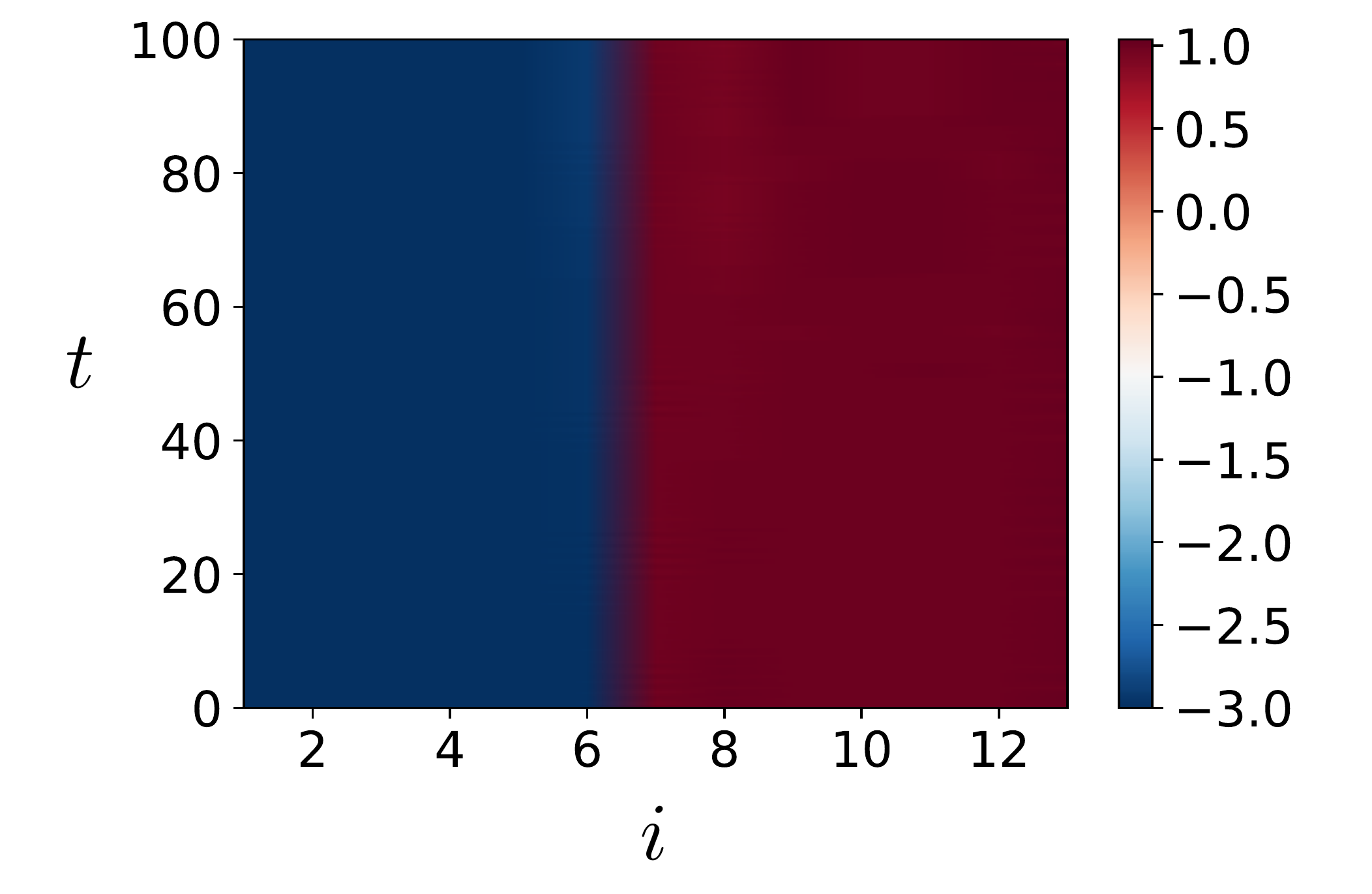}
\caption{
Energy density profile as a function of the coordinate $i$ along the chain [determined by exact diagonalization (ED) with $L=14$] and of time $t$, for $h_x = 2$, and $h_z =0.2$, where units are fixed such that $J=1$. In spite of the resulting resonance at the first order in $h_z$, no energy flow from right to left is observed.
}
\label{fig:resonance}
\end{figure}

\end{widetext}

\end{document}